\documentclass[twocolumn,showpacs,preprintnumbers,amsmath,amssymb]{revtex4}

\begin{document}

\title{Coping with the Pais-Uhlenbeck oscillator's ghosts in a canonical approach}

\author{Aldo D\'ector$^{1}$}
\email{dector@nucleares.unam.mx}
\author{Hugo A. Morales-T\'ecotl$^{1,2}$}
\email{hugo@xanum.uam.mx}
\author{Luis F. Urrutia$^{1}$}
\email{urrutia@nucleares.unam.mx}
\author{J. David Vergara$^{1}$}
\email{vergara@nucleares.unam.mx}

\affiliation{$^1$ Instituto de Ciencias Nucleares, Universidad Nacional Aut\'onoma de M\'exico,\\
A. Postal 70-543, M\'exico D.F., M\'exico,}

\affiliation{$^2$Departamento de F\'{\i}sica, Universidad Aut\'onoma Metropolitana Iztapalapa,\\
San Rafael Atlixco 186, C.P. 09340, M\'exico D.F., M\'exico,}

\date{\today}
\begin{abstract}
A {\em complex} canonical transformation is found that takes the fourth order derivative Pais-Uhlenbeck oscillator into two independent harmonic oscillators thus showing that this model has energy bounded from below, unitary time-evolution and no negative norm states, or ghosts.
Such transformation yields a positive definite inner product consistent with reality conditions in the Hilbert space.
The method is illustrated by eliminating the negative norm states in a complex oscillator.
Extensions to other higher order mechanical models and field theory are discussed.
\end{abstract}
\maketitle
{\em Introduction}.
Including higher derivatives in field theories was originally considered to improve their divergent ultraviolet behavior \cite{Thirring:1950}. Unfortunately their energy turned out not bounded from below \cite{Pais:1950za}, involving then ghosts, and making the theory non-unitary \cite{Heisenberg1957}. In spite of such sickness higher derivative theories were still studied to learn on their improved renormalization properties. Even a renormalizable higher order quantum gravity theory was advanced along these lines \cite{Stelle:1976gc} and the unitarity of a lattice form was studied in \cite{Tomboulis:1983sw} (see also \cite{Hawking:2001yt,Moeller:2002vx,Rivelles:2003jd, Antoniadis:2006pc,Codello:2006in,Berkovits:2006vc} and references there for other examples).
As for dealing with ghosts attempts can be classified according to whether the approach is perturbative \cite{Jaen:1986iz,Eliezer:1989cr,Cheng:2001du} or not \cite{Simon:1990ic,Hawking:2001yt,Rivelles:2003jd}, but a definite answer is yet to be found.

It is usually rewarding to study mechanical models instead of field theories to test new ideas and the higher-order derivative feature is not the exception. The model example to do so is the Pais-Uhlenbeck (PU) oscillator \cite{Pais:1950za} which consists of a one dimensional harmonic oscillator Lagrangian plus a term quadratic in acceleration.  A direct quantum treatment of the associated degrees of freedom gives rise to a spectrum not bounded from below since it  consists of the difference of two quantum harmonic oscillators spectra \cite{Pais:1950za}. An interesting solution to such difficulty has been recently proposed in \cite{Bender:2007wu} by finding a quantum transformation which gives its Hamiltonian a non-hermitian, $\cal PT$ symmetric form. In this approach a technique has been developed to obtain a physical inner product and thus a Hilbert space \cite{Bender:2007nj}. Some open problems with such a treatment, applied to the PU oscillator in \cite{Bender:2007wu}, include: (i) the hermitic properties of the final variables are not established in terms of the defined inner product,  (ii) an abnormal $\cal PT$ behavior is required for the intermediate canonical variables, (iii) the lack of a systematic approach to find the corresponding $\cal PT$ symmetric version for other higher-order derivative models and  (iv) its extension to field theories.

The aim of this Letter is to show that it is possible to tackle (i)-(iv) within a canonical approach. The latter is motivated by a previous proposal to build complex canonical variables for non-perturbative quantum canonical general relativity \cite{Ashtekar:1991hf,Ashtekar:1992ne}, which achieved the nontrivial task of making polynomial the constraints of general relativity whereas the so called reality conditions to be fulfilled by the non-hermitian variables were proposed to determine the inner product of Hilbert space. For canonical general relativity such reality conditions imply the metric of space and its rate of change  are real quantities. Some effort was devoted to exploit these complex variables (see for instance \cite{Thiemann:1995ug,Ashtekar:1995qw,Montesinos:1999qc}) however the use of real variables allowed finally for important progress \cite{Rovelli:2004tv,Thiemann:2007zz}.
For the sake of clarity we review below an example which yields an ordinary harmonic oscillator upon a complex canonical transformation \cite{Ashtekar:1992ne}. Although this is not a higher-order derivative problem it exhibits negative norm states. We will show below that the fourth order derivative Pais-Uhlenbeck model can be treated similarly. Unless otherwise stated we use units in which $\hbar=1$.

{\em The complex harmonic oscillator}.
Our analysis is much easily introduced by reviewing the example of a modified harmonic oscillator subject to a complex canonical transformation \cite{Ashtekar:1991hf,Ashtekar:1992ne}. Indeed this case has also been studied in a form that exploits the ${\cal PT}$ symmetry in the framework of the non-hermitian hamiltonian approach  \cite{Bender:2007nj,Jones,Swanson}.

Our starting point is the simple observation that the complex Lagrangian
\begin{eqnarray}\label{LC}
L_{C} &=& \frac{\dot{q}^2}{2}-\frac{q^2}{2}- i\epsilon q\dot{q}\,,
\end{eqnarray}
where $\epsilon$ is a real parameter, becomes the one corresponding to the harmonic oscillator after adding to it the total time derivative $\frac{df_{C}}{dt}, f_{C} = i\epsilon \frac{q^2}{2}$. In the Hamiltonian description the canonical momentum is
\begin{equation}
p = \frac{\partial L_C}{\partial \dot{q}}=\dot{q}-i\epsilon q\,,
\end{equation}
so  $p\in C\!\!\!\!I$ and the quantum Hamiltonian becomes

\begin{equation}
\label{Hmod}
\hat{H}_{C}=\frac{\hat{p}^{2}}{2}+\frac{\hat{q}^2}{2}-\frac{1}{2}\,\epsilon^{2}\hat{q}^{2}+\frac{i\epsilon}{2}\left\{\hat{p}\,,\hat{q}\right\}\,,
\end{equation}
where $\{\hat{p},\hat{q}\}=\hat{p}\hat{q}+\hat{q}\hat{p}$ so that a symmetric ordering is selected.
Upon the canonical transformation $\hat p=\hat P-i\epsilon \hat Q$, $\hat q=\hat Q$, with $\hat P=\frac{d\hat Q}{dt}$,  the Hamiltonian (\ref{Hmod}) becomes the standard one of the harmonic oscillator $\hat H_{HO}= \frac{\hat P^2}{2}+\frac{\hat Q^2}{2}$, which is hermitian whenever $\hat Q$ and $\hat P$ are.

The coordinate representation
\begin{equation}
\hat{q}=q\;, \qquad \hat{p}=-i\frac{d}{dq}\;,
\end{equation}
leads to a non-hermitian form of the Hamiltonian (\ref{Hmod}) with the usual scalar product \cite{rep}. The corresponding  Schr\"odinger equation becomes
\begin{equation}
\label{SchroHmod}
\psi''(\,q)-2\epsilon\, q\,\psi'(\,q)-\left((1-\epsilon^{2})q^2+\epsilon-2 E\right)\psi(\,q)=0\,.
\end{equation}
Hence the eigenvalue problem for (\ref{Hmod}) can be related to that of the harmonic oscillator by using the wave functions of the latter $\varphi_n=N_n\mathrm{e}^{-\frac{q^2}{2}}H_n(q)$ and defining
\begin{equation} \label{psin}
\psi_{n}(\,q):=e^{\frac{\epsilon q^{2}}{2}}\varphi_n\,,\quad E_{n}=n+\frac{1}{2}\,,\quad n=0,1,2\dots \,.
\end{equation}
A tedious but otherwise direct calculation shows that the eigenfunctions $\psi_n$  do not have positive norm in the Hilbert space ${\cal H}_0=L^2(I\!\!R,dq)$ \cite{DMTUV}. However, the classical role of $f_C$ as the generator of a canonical transformation motivates the introduction of a quantum similarity transformation such that (see for example \cite{Anderson:1993ia} for a general discussion)
\begin{equation}\label{CTho}
\hat{{H}}_{{HO}}=e^{-\epsilon \frac{q^2}{2}}\hat{{H}}_{C}\,e^{\epsilon \frac{q^2}{2}}\,.
\end{equation}
This non-unitary canonical transformation changes the measure $dq$ to $d\mu=\mathrm{e}^{-\epsilon q^2}\,dq $. Now the Hilbert space ${\cal H}=L^2(I\!\!R,d\mu)$ ensures (\ref{psin}) have positive norm since
\begin{eqnarray}
\langle n|m\rangle_{\mu}&:=&\int dq\, \mathrm{e}^{-\epsilon q^2}\psi_n^{\ast}\psi_m\nonumber\\
&=& \int dq\,\varphi_n^{\ast}\varphi_m
=: \langle n|m\rangle_1 =\delta_{mn}\,,
\end{eqnarray}
where subscripts denote the appropriate measure for the correspondent Hilbert space.

Moreover, the reality conditions $q^{\dag} = q\,,\quad  p^{\dag} = p + 2i\epsilon q$, induced by the canonical transformation, are implemented in $\cal H$ and make the Hamiltonian (\ref{Hmod}) hermitian \cite{Ashtekar:1991hf}.

Now, among the possible methods to deal with non hermitian variables the quantum action principle is particularly useful in regard to the dynamics \cite{Schwinger2001}. In what follows we use  $\hat p|p'\rangle= p'|p'\rangle$ and its adjoint form $\langle p''^{\ast}|\hat p^{\dagger}= \langle p''^{\ast}|p''^{\ast}$, to handle the non-hermitian  momentum. The resulting propagator takes the form
\begin{eqnarray} \label{prop}
\langle p^{\prime\ast},t &|& p^{\prime\prime},t=0\rangle =  A
\mathrm{e}^{i\frac{B(p^{\prime\prime 2}+ p^{\prime\ast 2})-2 p^{\prime\prime} p^{\prime\ast} }
{ C } }\,,\\
A&=& \frac{1}{\sqrt{2\pi i}}
\sqrt{\frac{1}{(\epsilon ^2+1)\sin(t) +2i\epsilon\cos(t)}} \,,\\
B&=& \cos(t)-i\epsilon\sin(t))\,,\\
C &=& 2((\epsilon^2+1) \sin(t)+2i\epsilon\cos(t))\,,
\end{eqnarray}
whose equal time and $\epsilon \rightarrow 0$ limit leads to $ \delta(p^{\prime\prime}-p^{\prime\ast})$, as it should be.
The bracket corresponding to the change of basis is readily shown to be
\begin{equation}
\langle P|p  \rangle = \sqrt{\frac{1}{2\pi\epsilon}} \mathrm{e}^{-\frac{1}{2\epsilon} \left( p-P \right)^2}\,,
\end{equation}                                                                                                                                                                            which allow us to relate the propagator expressed in terms of hermitian variables, $(\hat P,\hat Q)$, with that written in terms of non-hermitian ones, $(\hat p,\hat q)$, as follows
\begin{eqnarray}\label{PropagatorPp}
\langle p^{\ast},t_2| p',t_1\rangle =  \frac{1}{2\pi\epsilon} \int dPdP' &&\mathrm{e}^{-\frac{1}{2}((p^{\ast}-P')^2+(p'-P)^2)}
\nonumber \\
&& \langle P',t_2|P,t_1 \rangle\,.
\end{eqnarray}
The completeness relation in the non-hermitian description results
\begin{eqnarray}
\int d^2p\, \mu(p,p^*)\, |p\rangle\langle p^*| = 1\,,\quad
\mu =\frac{1}{\sqrt{\pi\epsilon}}\mathrm{e}^{+\frac{1}{4\epsilon}(p-p^*)^2}\,.
\end{eqnarray}
Thus far we can summarize the analysis of the model as follows. We provided a complex canonical transformation taking the complex harmonic oscillator (\ref{Hmod}) into the ordinary harmonic oscillator so that complex variables fulfill reality conditions in the Hilbert space ${\cal H}$ consistently with the canonical transformation. Neither ${\cal PT}$ symmetry nor an abnormal version of it has been invoked \cite{Bender:2007wu}. The states (\ref{psin}) possess positive definite norm in $\cal H$ and a unitary time evolution follows from the canonical transformation. Now we proceed with the higher order derivative model.

{\em The Pais-Uhlenbeck oscillator}. Let us start now from the PU oscillator Lagrangian
\begin{eqnarray} \label{LPU}
L_{\mathrm{PU}} &=& -\frac{1}{2}\ddot{x}^2 + \frac{(\omega_1^2 + \omega_2^2)}{2} \dot{x}^2 - \frac{\omega_1^2 \omega_2^2}{2} x^2\,,
\end{eqnarray}
where we assume from now on $\omega_1>\omega_2$. $L_{\mathrm{PU}}$ is connected to
\begin{equation}\label{Lxi}
L_{\xi} = \frac{1}{2} \dot{\xi_1}^2 - \frac{\omega_1^2}{2} \xi_1^2 +  \frac{1}{2} \dot{\xi_2}^2 - \frac{\omega_2^2}{2} \xi_2^2\,,
\end{equation}
for real $\xi_i, i=1,2$,  by the following relations
\begin{eqnarray}
L_{\mathrm{PU}} + \frac{df}{dt}&=& L_{\xi}\,,\label{LL}\\
f &=& -\dot{x}\ddot{x} \,,\\
\xi_1 &=& i (ax+b\ddot{x}) \,,\\
\xi_2 &=& cx + b\ddot{x} \,. \label{xi2}
\end{eqnarray}
By choosing $a,b,c$ to be real we see that $x$ is necessarily complex. To accomplish (\ref{LL}) the following values are obtained, for which the
same sign should be used,
\begin{eqnarray} \label{coeffPU}
\frac{a}{\omega_2^2} = b = \frac{c}{\omega_1^2} = \pm \frac{1}{\sqrt{\omega_1^2-\omega_2^2}} \,.
\end{eqnarray}
In other words, according to (\ref{LL}), $L_{\mathrm{PU}}$ in (\ref{LPU}) fails to be the real $L_{\xi}$ in (\ref{Lxi}) only by a time derivative. Note that $f$ will give rise to the corresponding canonical transformation once it is expressed in terms of phase space variables which we now derive. According to Ostrogradski's method applied to (\ref{LPU}) we get
\begin{eqnarray}
\Pi_x &=& (\omega_1^2 +\omega_2^2) \dot{x} + \stackrel{...}{x} \,,\\
z &=& \dot{x} \,,\\
\Pi_z &=& - \ddot{x} \,,
\end{eqnarray}
so that $\{x,\Pi_x\}=1$ and $\{z,\Pi_z\}=1$. Clearly $f=z\Pi_z$. The quantum Hamiltonian is
\begin{equation}\label{HPU}
\hat H_{\mathrm{PU}} = - \frac{1}{2} \hat \Pi_z^2 - \frac{\omega_1^2 + \omega_2^2}{2} \hat z^2 + \frac{1}{2}\left\{\hat z,\hat \Pi_x\right\} + \frac{\omega_1^2 \omega_2^2}{2} \hat x^2 \,.
\end{equation}
Using the afore mentioned transformation we have
\begin{eqnarray}
\hat x &=& ib\hat \xi_1 + b\hat \xi_2 \,,\label{xxi1}\\
\hat \Pi_x &=& ia \hat P_1 + c \hat P_2 \,, \label{xxi2}\\
\hat z &=& ib \hat P_1 + b \hat P_2 \,,\label{xxi3}\\
\hat \Pi_z &=& ic \hat \xi_1 + a \hat \xi_2\,, \label{xxi4}
\end{eqnarray}
where $\hat P_i, i=1,2$, are the canonical momenta corresponding to $\hat \xi_i$ obtained from (\ref{Lxi}). In terms of these hermitian variables the hamiltonian (\ref{HPU}) takes the form
\begin{equation}\label{HHO}
\hat H_{\xi} = \frac{\hat P_1^2}{2} + \frac{\omega_1^2}{2} \hat \xi_1^2 + \frac{\hat P_2^2}{2} + \frac{\omega_2^2}{2}\hat  \xi_2^2 \,.
\end{equation}
So, starting from a Hamiltonian that is not bounded from below a well defined Hamiltonian in terms of $P_i$ and $\xi_i$ is obtained.
Notice that in \cite{Bender:2007wu} it is acknowledged there exists a similarity transformation relating the original PU oscillator with a couple of independent harmonic oscillators similar to our case. Nevertheless a further transformation has to be supplied in \cite{Bender:2007wu} to obtain the right frequencies for the oscillators \cite{Anderson:1993im}.

The propagator for the PU model obtained by Schwinger's formalism takes the form
\begin{eqnarray} \label{PropPU}
 && \langle x^{\prime*}, \Pi_z^{\prime*},t | x, \Pi_z,t=0\rangle = \left. \exp\left\{ \frac{i}{D}\right\{F(x^{\prime*}\Pi_z+x\Pi_z^{\prime*})
  \right.\nonumber\\
  && +G \Pi_z\Pi_z^{\prime*} +J \left(x^{\prime*}\Pi_z^{\prime*}+x\Pi_z\right)+K\left(\frac{\Pi_z^{\prime*2}+\Pi_z^{2}}{2}\right)\nonumber\\
   && \left.\left. + M x x^{\prime*}+ N \left(\frac{x^{\prime*2}+x^{2}}{2}\right)\right\}\right\}\,,
\end{eqnarray}
with $D,F,G,J,K,M,N$, being the following functions of $t$:
\begin{eqnarray}
  D &=& (\omega_1^2-\omega_2^2)\sin(\omega_1t)\sin(\omega_2t), \nonumber\\
  F &=&  \omega_1\sin(\omega_1t)+\omega_2\sin(\omega_2t),\nonumber\\
  G &=& -\omega_2\sin(\omega_1t)-\omega_1\sin(\omega_2t), \nonumber\\
  J &=& \!\!-\omega_1\sin(\omega_1t)\cos(\omega_2t)+\!\! \omega_2 \sin(\omega_2t)\cos(\omega_1t),\nonumber\\
  K &=& \!\! \omega_2\sin(\omega_1t)\cos(\omega_2t)-\!\! \omega_1 \sin(\omega_2t)\cos(\omega_1t) ,\nonumber\\
  M &=& -\omega_1^3\sin(\omega_1t)-\omega_2^3\sin(\omega_2t), \nonumber\\
  N &=&\omega_1^3\sin(\omega_1t)\cos(\omega_2t)-\!\! \omega_2^3
  \sin(\omega_2t)\cos(\omega_1t)\,.
\end{eqnarray}
Just as in the previous case of a complex harmonic oscillator the PU propagator (\ref{PropPU}) can be related to the one corresponding to the
hamiltonian (\ref{HHO}) \cite{DMTUV} by using the change of basis
\begin{equation}\label{coneco}
    \langle P_{1},P_{2}|x, \Pi_z\rangle = \exp\left[(ax-b\Pi_z)P_{1}+(-icx+ib\Pi_z)P_{2}\right]\,.
\end{equation}
The basis $|x,\Pi_z\rangle$ is complete with the measure
\begin{equation}\label{meas}
    d\mu_{_{PU}}=\frac{d\alpha d\beta d\gamma d\varrho}{(2\pi)^2}\delta(b\gamma-a\alpha)\delta(b\varrho-c\beta),
\end{equation}
Here $x=\alpha +i \beta $ and $\Pi_z=\gamma +i\varrho$ where $\alpha,\beta,\gamma,\varrho$ are real.

{\em Discussion}.
In this Letter we have advanced a description of the four-order derivative PU oscillator (\ref{LPU}) based on a complex canonical transformation (\ref{xxi1})-(\ref{xxi4}) that connects it with two decoupled harmonic oscillators (\ref{HHO}) with just the corresponding frequencies $\omega_1$ and $\omega_2$, appearing in (\ref{LPU}). This shows that the PU oscillator is ghost-free,  has energy bounded from below and its time evolution is unitary.
Our treatment was motivated by the use of complex variables for non-perturbative canonical quantum general relativity \cite{Ashtekar:1991hf,Ashtekar:1992ne} that simplified form of its constraints.  The reality conditions were implemented in the Hilbert space. A simpler model leading to an ordinary  harmonic oscillator was first described to illustrate the features of the complex canonical transformation. Finally the method was extended to the truly higher-order derivative PU oscillator.

We propose to generalize our approach to other mechanical quadratic higher-order derivative models by following (\ref{LL})-(\ref{xi2}).
Let us consider a system whose Lagrangian contains up to the $n$-th order time derivative in a quadratic form. Such a Lagrangian, $L(x,\dot{x},\dots,x^{(n)})$, where $x^{(k)}= \frac{d^k x}{dt^k}$, yields an equation of motion of order $(2n)$
\begin{eqnarray}
\sum_{k=0}^{n} (-1)^k\frac{d^k\mbox{}}{dt^k} \frac{\partial L}{\partial x^{(k)}} &=&0\,.
\end{eqnarray}
Assuming $x$ and hence $L$ to be real leads to the well known problems of the higher-order derivative models. However as shown in this work the introduction of complex (non-hermitian) variables leads to their resolution. Specifically we assume the complex description can be related to another in terms of a real Lagrangian $L_{\xi}(\xi_1,\dot{\xi}_1,\dots,\xi_n,\dot{\xi}_n)$ through
\begin{eqnarray}
L(x,\dot{x},\dots,x^{(n)}) &+& \frac{df}{dt} = L_{\xi}(\xi_1,\dot{\xi}_1,\dots,\xi_n,\dot{\xi}_n) \,,\\
f &=& f(x,\dots,x^{(n)})\,\\
\xi_i &=& \xi_i(x,\dots,x^{(n)}), i=1,\dots, n \,.
\end{eqnarray}
Here $\xi_i, i=1,\dots,n$ are real coordinates and only their first order time derivatives $\dot{\xi}_i$ in a quadratic form enter in $L_{\xi}$. In general $f$ will be complex. The $\xi_i$ select the real sector of the complexified model. A complete systematic analysis is obviously required and it is currently in progress \cite{DMTUV}.

As for the case of a quadratic higher-order derivative free scalar field model our approach works similarly. The following Lagrangian density
\begin{equation}
{\cal L} = -\frac{1}{2} (\Box \phi)^2 + \frac{m_1^2+m_2^2}{2} \partial_\mu\phi\partial^\mu\phi -\frac{m_1^2 m_2^2}{2} \phi^2 \,,
\end{equation}
is related to
\begin{equation}\label{Lpsi}
{\cal L}_{\psi} = \frac{1}{2} \partial_\mu\psi_1 \partial^\mu \psi_1 -\frac{m_1^2}{2} \psi_1^2 +\frac{1}{2} \partial_\mu\psi_2 \partial^\mu \psi_2 -\frac{m_2^2}{2} \psi_2^2 \,,
\end{equation}
by
\begin{eqnarray}
{\cal L} + \partial_{\mu} f^{\mu}&=& {\cal L}_{\psi}\,,\\
f^{\mu} &=& - \Box \phi \partial^{\mu}\phi  \,,\\
\psi_1 &=& i(a\phi  + b\Box \phi) \,,\\
\psi_2 &=& c\phi  + b\Box \phi \,,
\end{eqnarray}
with $a,b,c$ having the same form as in the PU model (\ref{coeffPU}) except for the replacing $\omega_i\rightarrow m_i , i=1,2$.
Clearly (\ref{Lpsi}) is ghost-free. Thus our approach looks promising \cite{DMTUV} even in the interacting case including a term of the type
$(\phi\phi^{\ast})^2= -\frac{1}{(c-a)^4}(\psi_1^2+\psi_2^2)^2$ \cite{Hawking:2001yt,Antoniadis:2006pc}.

Our analysis of the PU oscillator is an alternative to that of  \cite{Bender:2007wu} based on  non-hermitian but ${\cal PT}$ symmetric Hamiltonians. We consider that our method has the following advantages: 1) it is based on non-hermitian variables satisfying reality conditions arising from (\ref{xxi1})-(\ref{xxi4}) that make the higher-order Hamiltonian hermitian in the appropriate Hilbert space with  measure (\ref{meas}), 2) no $\cal PT$ symmetry is required, 3) it can be translated into a free higher-order field theory without difficulty. Although promising the interacting case requires further study.  And finally 4) it seems possible to extend the method at least to quadratic higher-order derivative mechanical models.

{\em Acknowledgments} This work was partially supported by Mexico's National Council of Science
 and Technology (CONACyT), under grants CONACyT-SEP 51132F, CONACyT-SEP 47211-F,  CONACyT-SEP 55310, DGAPA-UNAM IN109107 and a CONACyT sabbatical grant to HAMT. AD wishes also to acknowledge  support from CONACyT.


\begin{thebibliography}{99}
\bibitem{rep}
Since the realization of the momentum is defined up to function of $q$, using $\hat{p}=-i\frac{d\mbox{}}{dq} + i\epsilon q$ would yield the usual hamiltonian for the harmonic oscillator which would however require a modified inner product to be hermitian.
\end{thebibliography}
\end{document}